\documentclass[preprint,review,3p,12pt]{elsarticle}

\usepackage{amssymb}
\usepackage{amsmath}

\usepackage{graphicx}
\usepackage{epstopdf}
\usepackage{subfig,float}
\usepackage{natbib}
\usepackage[dvipsnames]{xcolor}
\usepackage{dcolumn}
\usepackage{bm}
\newcommand{\lra}[1]{\langle #1 \rangle }

\def\be{\begin{equation}}
\def\ee{\end{equation}}

\journal{International Journal of Multiphase Flow}

\begin{document}

\begin{frontmatter}

\title{Assessment of the point-wise approach for the Turbulent Settling  of finite-size  particles} 

\author[1]{Francesco Battista\corref{cor1}%
\fnref{fn1}}
\ead{francesco.battista@uniroma1.it}
\address[1]{Department of Mechanical and Aerospace Engineering. Sapienza University, via Eudossiana 18, 00184, Rome, Italy}

\author[2]{Sergio Chibbaro%
\fnref{fn2}}
\ead{sergio.chibbaro@universite-paris-saclay.fr}
\address[2]{Universit\'e Paris-Saclay, CNRS, UMR 9015, LISN, F-91405, Orsay Cedex,France}

\author[1]{Paolo Gualtieri%
\fnref{fn1}}
\ead{paolo.gualtieri@uniroma1.it}

\cortext[cor1]{Corresponding author}




\begin{abstract}
We study the settling of suspensions of relatively large particles with a diameter of the order of ten Kolmogorov scales and density slightly larger than the carrier fluid in statistically steady homogeneous isotropic turbulence. 
The particle-to-fluid density ratio is varied to obtain a wide range of Galileo numbers, which are the ratios between buoyancy and viscous forces. 
We analyse the problem through high-resolved one-way coupled direct numerical simulations where the particles are modeled as material points. The physical parameters are chosen in the same range used in recent particle-resolved simulations (PRS,~\citep{fornari2016sedimentation,fornari2016reduced}), against which we compare. 
The results of the point-wise simulations are in good agreement with the PRS ones, showing a reduced settling speed for the range of parameters under investigation, relevant to suspensions settling in aqueous media, at volume fractions up to a few percent for density ratios order of one. Results are obtained neglecting the inter-particles and particle-fluid interactions while purposely including/not including the different forces (e.g. Stokes drag, added mass, lift force) in the particles' equations of motion to highlight their contributions respectively.
At a high Galileo number, the mean settling velocity is only slightly affected by turbulent fluctuations, and it is the same obtained for the settling velocity of a single particle in a quiescent fluid. When the Galileo number is reduced, the settling velocity is progressively affected by turbulent fluctuations that cause a substantial decrease in the particle sedimentation speed. 
The transition occurs in a range of parameters where the settling velocity of the isolated particle in a quiescent fluid is the order of the root mean square value of the turbulent fluctuations. 
The present results are particularly relevant for applications. Point-wise models endowed with an accurate description of the hydrodynamic force are effective in capturing the particle settling speed and other higher-order 
statistics as demonstrated by direct comparison against particle-resolved simulations~\citep{fornari2016sedimentation,fornari2016reduced}.
\end{abstract}



\end{frontmatter}

%

\section{Introduction}

Inertial particles are present in fluid flows in many engineering and environmental applications~\cite{forterre2008flows,balachandar2010turbulent,guazzelli2011physical,andreotti2013granular,elghobashi2019direct,brandt2022particle}. 
Among the most relevant, we can mention the transport of sediments, particle dispersion, granular flows, and volcanic eruptions. 
A key ingredient of all those processes is particle settling~\cite{fokeer2004characterisation,noh2006large,keshtpoor2015numerical}, on which we focus in the present work.

The dynamics depend on several physical quantities entering the experimental conditions, which may be represented by dimensionless parameters~\cite{clift2005bubbles}.
In particular, the particle-to-fluid density ratio $\rho_p/\rho_f$, the volume fraction of the suspension $\phi_V$, and the ratio between the buoyancy and the viscous forces given by the Galileo number $Ga=\sqrt{(\rho_p/\rho_f-1)g d^3 / \nu^2}$, where $g$ is the acceleration of gravity, $d$ the diameter of the particle, and $\nu$ the kinematic viscosity. Furthermore, settling may occur in a quiescent fluid or a turbulent flow. In turbulent flows, velocity fluctuations may significantly affect the settling velocity. In this case, it is also necessary to consider the Stokes number (the ratio between the particle and the flow time scales) and the Reynolds number. In isotropic conditions, it is practical to use the Taylor-scale Reynolds number $Re_\lambda=u_{rms} \lambda /\nu$, where $u_{rms}$ is the root-mean-square of the velocity fluctuations and $\lambda$ is the Taylor length-scale~\cite{pope2000turbulent}.

Despite many studies, the impact of turbulence on settling remains somewhat unclear. Given the several dimensionless parameters that control the phenomenon, intensive research is ongoing.
Following an analysis of a simplified flow~\cite{maxey1986gravitational,maxey1987gravitational}, 
many Direct Numerical Simulations (DNS) and experiments provide evidence of an increase of the 
settling speed for small inertial heavy particles~\cite{wang1993settling,nielsen1993turbulence,good2014settling,rosa2016settling}, i.e. for particles with density ratios much larger than one and particle diameter smaller than the Kolmogorov scale. Enhancement of the settling velocity has also been found in detailed studies of laminar and turbulent 2D flows~\cite{de2008sedimentation,afonso2008terminal}.
The dominant mechanism is a preferential sweeping leading to a bias in the trajectories.
Other possible mechanisms for the settling enhancement may be the particle-particle hydrodynamic interactions~\cite{aliseda2002effect} and two-way coupling effects~\cite{bosse2006small}, both becoming important when increasing the volume fraction.

Some of the same studies and other ones suggest that turbulence may inhibit settling when the control parameters are in specific ranges; for instance, when considering almost neutrally buoyant particles (particle-to-fluid density ratio order one) or relatively large particles (Stokes numbers order ten) ~\cite{nielsen1993turbulence,yang2003settling,kawanisi2008turbulent,afonso2008terminal,good2014settling}.
Sometimes, controversial results have been reported. In particular, it has been claimed that the phenomenon of the reduction of the settling velocity may be observed in numerical simulations of heavy particles only when a nonlinear drag is used~\cite{good2014settling}, whereas a subsequent exhaustive study~\cite{rosa2016settling} reports always enhancement of the settling velocity both with linear and nonlinear drag.
These studies underline that the problem of settling is rather complex and depends on many 
independent parameters. Moreover, results are sensitive to the large-scale flow structures, and particular attention has to be paid to the set-up and the subsequent numerical simulation~\cite{rosa2016settling}.

The above studies focus on small heavy particles with an associate particle Reynolds number of order one. 
In this framework, only viscous drag and gravity play a role, and nonlinear corrections are negligible~\cite{rosa2016settling}. However, in relevant geophysical and hydraulic applications, water serves as the carrier fluid. Hence, for common materials, the particle-to-fluid density ratio is of order one, and the particle diameter is slightly larger than the Kolmogorov scale. 
In this range of parameters, all the different contributions to the hydrodynamic 
forces on the particles should be retrieved. However, the expression of the force is based on an 
asymptotic expansion~\cite{gatignol1983faxen,maxey1983equation} that may be a source of errors when particles have a diameter of the order of the Kolmogorov scale or larger, or the Reynolds number associated with particles is large. As seen above, the diameter of the particles is an important feature since it enters the volume fraction and the Galileo number, which controls the buoyancy effects.
Moreover, for the sake of practicality, most of the studies on the subject have dealt with one-way coupling, neglecting the effects of the particles on the fluid. The one-way coupling 
point-wise approach is safe for particles smaller than the Kolmogorov scale and in dilute conditions~\cite{toschi2009lagrangian,elghobashi2019direct}. Yet, when the number of particles is high and/or the diameter is large, the flow becomes semi-dilute or even moderately dense, with a dramatic impact on the flow~\cite{forterre2008flows,gualtieri2013clustering,capecelatro2014numerical,elghobashi2019direct,marchetti2021falling}.
On the other hand, even in dilute conditions, the interaction between particles larger than the Kolmogorov scale and the flow may lead to non-trivial phenomena, including long-range interactions, e.g. wakes. 
In these cases, both the point-wise and the one-way coupling approach may fail, and particle-resolved simulations (PRS) should be used~\cite{uhlmann2008interface}, or at least consistent two-way coupling effects should be added~\cite{gualtieri2015exact}.

The purpose of the present work is to assess to what extent point-wise, one-way coupled simulations may be used in a region of the parameter space different from that usually studied, namely for particles with a diameter larger than the Kolmogorov scale, yet at a density ratio of order one.
 
In the last years, several studies have investigated finite-size effects, looking at the settling of inertial particles in quiescent conditions~\cite{yin2007hindered,uhlmann2014sedimentation} or at neutrally buoyant particles in isotropic turbulence~\cite{homann2010finite,byron2015shape}.
More relevant for the present study, in~\cite{fornari2016sedimentation,fornari2016reduced}, the authors have carried out several high-fidelity particle-resolved direct numerical simulations to compare quiescent to turbulent settling velocity and to analyse the effect of Galileo's number in a turbulent environment for particle-to-fluid density ratio of order one. Those results confirmed some trends pointed out in previous experiments~\cite{yang2003settling}.
These studies show a clear reduction of the particle settling velocity due to the background 
turbulent flow. Moreover, a decrease in Galileo's number may strongly enhance this effect, at variance with what was found for small heavy particles~\cite{wang1993settling,good2014settling,rosa2016settling}.

The present work aims to compare such predictions with direct numerical simulations of point-wise particles in a one-way coupling regime. The goal is to understand whether such a drastically simplified approach, which requires much less computational burden, may be used in place of the full-resolved particle simulations in some range of the parameters. 
Indeed, this would offer a valuable alternative for engineering applications and provide insights into the settling process.
 
Interestingly, and somehow surprisingly, the present results show in the range of parameters chosen in the fully-resolved simulations~\cite{fornari2016sedimentation,fornari2016reduced}, that the settling velocity is well captured by point-wise dynamics when all the relevant hydrodynamic forces are included in the dynamics of the particles.

\section{Model Equations and dimensional analysis}

Direct Numerical Simulations of homogeneous and isotropic turbulence are performed in 
a $(2\pi)^3$ periodic box by solving the dimensionless Navier-Stokes equations,
\begin{equation}
\label{eqn:ns_hi}
\left\{
\begin{array}{ll}
\displaystyle \nabla \cdot {\bf u} = 0 \\ 
\displaystyle \frac{\partial {\bf u}}{\partial t} +  {\bf u}  \cdot \nabla  {\bf u} 
 = - \nabla p + \frac{1}{{\rm Re}} \nabla^2 {\bf u} + {\bf f}  \, ,
\end{array}
\right.
\end{equation}
where $Re=u_0 \ell_0/\nu$ is the  {\it nominal} Reynolds number. Equations are made 
dimensionless using arbitrary {\it nominal} velocity and spatial reference scales $u_0$ and 
$\l_0$ respectively, and $\nu$ is the kinematic viscosity. In these units, the 
Reynolds number should be interpreted as the inverse of the dimensionless viscosity
$\bar{\nu}=1/Re$ and is the only dimensionless parameter in the Navier-Stokes equations. 
Therefore, $u_0$, $\l_0$, and $\nu$ do not need to be explicitly given, but only $Re$ matters.
The dimensionless external field ${\bf f} ({\bf x},t)$ is a large-scale random 
forcing~\cite{alvelius1999random} that allows reaching statistically steady conditions and
allows to set {\em a priori} the dimensionless mean energy dissipation rate of the flow
that statistically equals the power input.
Two parameters, therefore, characterise the fluid phase, $Re$ and the amplitude of the forcing.
Namely, all the simulations are performed with a dimensionless unitary power input or equivalently
dimensionless dissipation $\bar{\epsilon} = \epsilon /(u_0^3/\ell_0)= 1$, and $Re=200$. 
Only {\it a posteriori}, we can compute the customary Taylor-based Reynolds number 
$Re_\lambda= u_{rms} \lambda / \nu = 100$.

The dimensionless Basset-Boussinesq-Oseen equations describe the dynamics of point-wise particles~\cite{gatignol1983faxen,maxey1983equation},
\begin{equation}
\label{eqn:prtcls}
\left\{
\begin{array}{ll}
\displaystyle \frac{d {\bf x}_p}{dt}  &=  {\bf v}_p(t)  \\  \\
\displaystyle \frac{d {\bf v}_p}{dt}  &=  \dfrac{1}{St} \left( {\bf u}|_p - {\bf v}_p +
\frac{d_p^2}{24} \nabla^2 {\bf u}|_p  \right) +
\displaystyle \dfrac{3/2}{\left( \rho_p / \rho_f + 1/2 \right)}  \frac{D {\bf u}}{Dt}\bigg|_p \\
 &\displaystyle + \frac{\rho_p / \rho_f - 1}{Fr^2 \left( \rho_p / \rho_f + 1/2 \right)} {\bf \hat{e}_g} +
\displaystyle \frac{1/2}{\left( \rho_p / \rho_f + 1/2 \right)} \left( {\bf u}|_p - {\bf v}_p \right) \times \boldsymbol{\omega}|_p \, 
\end{array}
\right.
\end{equation}
where all the forces acting on a spherical particle, except the Basset history term, have been 
considered, namely viscous drag,  unsteady pressure gradient, and added mass terms, including Fax\'en corrections~\cite{gatignol1983faxen,maxey1983equation,gatignol2007history}, {and the Auton's lift force~\cite{auton1987lift} that calls for the evaluation of the fluid vorticity $\boldsymbol{\omega}$
at the particle position.}

In eq. \eqref{eqn:prtcls}, $St=\tau_p / \tau_0$ is the reference Stokes number where $\tau_p$ is 
the particle time-scale $\tau_p = (1/18 \nu) (\rho_p / \rho_f + 1/2) {d}^2$ with $\tau_0=\ell_0/u_0$, 
and can be expressed as $St=(Re/18) (\rho_p / \rho_f + 1/2) d_p^2$,  where $d_p$ is the 
dimensionless particle diameter, i.e. $d_p={d}/\ell_0$. $Fr= u_0 /\sqrt{g \ell_0} $ 
is the Froude number and  ${\bf \hat{e}_g}$ is the unit vector in the direction of the gravity 
acceleration ${\bf g}$. 

Equations \eqref{eqn:prtcls} are obtained by evaluating the fluid stress on a spherical particle. 
The hypotheses supporting their validity are: i) $Re_p= w \, d  /\nu \ll 1$ where $w$ is a representative 
scale for the particle-to-fluid relative velocity ${\bf w}={\bf u}|_p - {\bf v}_p$; 
ii) $(d^2/\nu)/(u_0/\ell_0)\ll1$, which gives the ratio between the particle viscous time and a typical 
flow time-scale; iii) $d/\ell_0\ll1$. It is worth underlying that even though Fax\'en correction takes into account finite-size effects, i.e. the presence of a steady but non-uniform flow field far from the particles, this correction is appropriate only in the limit of small diameter $d/\ell_0\ll1$. It is known that for moderately large particles, the correction is insufficient, and higher order terms are important~\cite{happel2012low,batchelor1972determination}.

In the case of heavy particles, $\rho_p/\rho_f \gg 1$, equations \eqref{eqn:prtcls} 
can be simplified into $ {d {\bf v}_p}/{dt}  =  {1}/{St} \left( {\bf u}|_p - {\bf v}_p \right) + 1/Fr^2 {\bf \hat{e}_g}$. 
However, for particle-to-fluid density ratios order one, the other terms may be key~\cite{clift2005bubbles}.
It is important to note that we neglect the Basset history force, as done in almost all numerical point-wise studies~\cite{balachandar2010turbulent,brandt2022particle}, although that is not apriori justified, and the history term has been found to provide a significant contribution in many respects~\cite{bergougnoux2014motion,olivieri2014effect,daitche2015role}.

For heavy particles, it has been proposed on purely heuristic grounds a method to generalise the particle equation of motion to the case when $Re_p \gg 1$. In this case, it has been suggested~\cite{clift2005bubbles} to correct the viscous drag nonlinearly, with many empirical formulas available.
One of the most used corrections is given by the Schiller-Naumann formula, which is the drag coefficient $C_D(Re_p)=24/Re_p \left( 1 + \alpha Re_p^\beta \right)$, where the coefficients are 
$\alpha=0.15$ and $\beta=0.687$. This formula gives a satisfactory agreement with experiments of 
settling of heavy particles in a quiescent fluid for a large range of the Galileo (or equivalently Froude) 
number, and notably in the range of parameters used here~\cite{clift2005bubbles}. With the addition of this correction, the equations of motion become:
\begin{equation}
\label{eqn:prtcls2}
\left\{
\begin{array}{ll}
\displaystyle \frac{d {\bf x}_p}{dt}  =  {\bf v}_p(t)  \\ \\
\displaystyle \frac{d {\bf v}_p}{dt}  =  \frac{1}{St} \left[ (1 + \alpha Re_p^\beta) ({\bf u}|_p - {\bf v}_p ) +
\frac{d_p^2}{24} \nabla^2 {\bf u}|_p  \right] +
\frac{3/2}{\left( \rho_p / \rho_f + 1/2 \right)}  \frac{D {\bf u}}{Dt}\bigg|_p \\ 
~~~~~~\displaystyle +
\frac{\rho_p / \rho_f - 1}{Fr^2 \left( \rho_p / \rho_f + 1/2 \right)} {\bf \hat{e}_g} 
 + \frac{1/2}{\left( \rho_p / \rho_f + 1/2 \right)} \left( {\bf u}|_p - {\bf v}_p \right) \times \boldsymbol{\omega}|_p\, .
\end{array}
\right.
\end{equation}

However, when the density ratio is of order one, other phenomena like secondary motion and slip may be important~\cite{clift2005bubbles}. In this case, the heuristic formula might not be satisfactory, and it might be necessary to tune different terms. One of the objectives of the work is to investigate this issue.

In the present framework, dimensional analysis shows that the problem is controlled by four 
dimensionless parameters, including the one related to the turbulent fluid motion, $Re$. 
In contrast, the volume fraction does not enter the problem since one-way coupling is considered. 
The particles' terminal velocity can then be expressed as
$v_t/u_0=\phi(\rho_p/\rho_f, Fr, St, Re)$, with $\phi$ some universal function.
In principle, a second turbulence parameter could enter since turbulence is forced. 
However, since the forcing is fixed, we omitted the 5th parameter.

As anticipated in the introduction, other choices for the parameters are, however, possible. In settling problems, it is 
usually considered the Galileo number $Ga=\sqrt{(\rho_p/\rho_f-1) \, g \,d^3/\nu^2 }$ that can be expressed in terms 
of the original  parameters as \\
$Ga=(Re/Fr) \sqrt{\rho_p/\rho_f-1} \left[18 St / (\rho_p/\rho_f+1/2) \right]^{3/4}$, 
and the ratio between the 
particle diameter and the Kolmogorov scale $d/\eta$ that can be expressed as 
$d/\eta=\sqrt{18 St_\eta / (\rho_p/\rho_f +1/2)}$ where $St_\eta=\tau_p / \tau_{\eta}$ and $\tau_{\eta}$ is
the Kolmogorov time scale. Note that $St_\eta=St/( \bar{\eta}^2 Re)$ where $\bar{\eta}$ is the Kolmogorov
scale in computational units $\bar{\eta}= \eta/\ell_0=\left(\bar{\nu}^3 / \bar{\epsilon} \right)^{1/4}$.
Hence, the dependence of the (dimensionless) settling velocity might equivalently be written as
$v_t/u_0=\phi(\rho_p/\rho_f, Ga, d/\eta, Re_\lambda)$.

In the present work, we compare with the particle-resolved simulations discussed 
in \cite{fornari2016sedimentation,fornari2016reduced} where particles with diameter $d / \eta = 12$ and different Galileo's numbers settling in homogeneous isotropic turbulence at $Re_\lambda = 90$ were considered.
 The Galileo number is changed by adjusting the particle-to-fluid density ratio $\rho_p/\rho_f$.
The equations (\ref{eqn:prtcls2}) are numerically solved in our code.
As said before, given  $\rho_p/\rho_f$, $Ga$, $d / \eta$, and $Re_\lambda$,
all the dimensionless parameters in eq. \eqref{eqn:prtcls2} can be computed at once.

In Table \ref{tab:HI_tab}, we report the other dimensionless parameters characterising
the turbulent flow. From a numerical point of view, the flow is resolved by using a
pseudo-spectral code with $N=256^3$ Fourier modes endowed with the $3/2$ dealiasing procedure
when computing the nonlinear terms in physical space. This provides a resolution of $k_{max}\eta=4.17$ (where $k_{max}=\sqrt{3}N/2$), which fulfils the requirements reported in the literature.  Equations \eqref{eqn:ns_hi} and \eqref{eqn:prtcls2}
are integrated in time by a four-stage, third-order, low-storage Runge-Kutta method.
The present numerical approach has been extensively used and validated in several 
configurations in previous studies~\cite{motta2020application}. We refer to them for more 
numerical details.

Concerning the dispersed phase, the particles are initialized with a random homogeneous 
spatial distribution, and their initial velocity matches the fluid velocity.  An initial transient 
corresponding to 10 $\tau_p$ is neglected before statistics are collected. To achieve 
a good statistical convergence $10^6$ particles are followed in the turbulent flow for each run.

\begin{table}
\begin{centering}
\begin{tabular}{cccccc}
 $Re_\lambda$    & $u_{rms}/u_\eta$ & $L_0/\eta$ &   $(St_\eta)_{min/max}$  & $St_{min/max}$ & $d/\eta$ \\  
  \hline
 100                     &  4.82                    & 150                 & 11.80/12.58           &  0.85/0.90 &12    \\
 \end{tabular}
\caption{Fluid and particle parameters. $L_0=k^{3/2}/\bar{\epsilon}$ is the integral scale where $k$ is 
the turbulent kinetic energy. $u_{rms}$ is the root mean square of the velocity fluctuations and
$u_\eta$ is the Kolmogorov velocity scale. Different $Ga$ number populations are considered, namely 
$Ga={1,5, 10, 20, 60, 145, 200}$. The Galileo number is adjusted by changing the density 
ratio $\rho_p/\rho_f$. } 
\label{tab:HI_tab}
\end{centering}
\end{table}
\begin{table}[b]
\begin{centering}
\begin{tabular}{c|ccccccc}

 $Ga$                                    & 1 & 5 & 10 & 20 & 60  & 145 & 200 \\ \hline
 data ~\cite{clift2005bubbles}& 0.0012 & 0.0264 & 0.0888& 0.2691& 1.2891& 4.0645 & 6.0892 \\
 DNS data                             & 0.0013 & 0.0262 & 0.0889& 0.2693& 1.2890& 4.0642 & 6.0892 
  \end{tabular}
\caption{Dimensionless terminal velocity in a quiescent fluid as computed from \cite{clift2005bubbles} 
and from present DNS data for different $Ga$ numbers.
\label{tab:terminal_tab}}
\end{centering}
\end{table}

\section{Results}

The balance between the buoyancy force and the global hydrodynamic force $F_D$
for a spherical particle settling in a quiescent fluid leads to the general formula for the terminal 
Reynolds number $Re_t=v_t d /\nu$: 
\begin{equation}
C_D Re_t^2 = \dfrac{4 \rho_f \Delta \rho g d^3}{3 \mu^2}= \frac{4}{3}Ga^2~,
\end{equation}
where the drag coefficient is $C_D\equiv 2 F_D/\rho_f \,v_t^2 \,S$,  $v_t$ is the 
terminal velocity of the isolated particle in a quiescent fluid, $S=\pi d^2/4$ is 
the area of the frontal section of a sphere, and $\Delta \rho = \rho_p - \rho_f$.
In the Stokes regime since $C_D = 24/Re_t$, the terminal Reynolds number $Re_t$ is proportional to 
$Ga^2$. That is equivalent to the scaling  
$v_t \propto \Delta \rho \,g \ d^2 /\mu$, i.e. the terminal velocity scales
as the square of the particle diameter. In the opposite regime of large particle terminal 
Reynolds numbers, say for $Re_t > 10^3$, the Drag coefficient approaches a plateau before the 
Drag crisis occurs at $Re_t \simeq 4 \cdot 10^5$. In this range, $Re_t \propto Ga$, that 
is equivalent to $v_t \propto \sqrt{\Delta \rho \,g \ d}$, i.e. the terminal velocity scales 
as the square root of the particle diameter. 

For the cases considered here, we show in the Table \ref{tab:terminal_tab} the terminal 
velocity of a single particle in a quiescent fluid comparing that taken from the 
literature~\cite{clift2005bubbles}, with those from the present numerical simulations.

\begin{figure}[b!]
\centering
\includegraphics[width=0.49\textwidth]{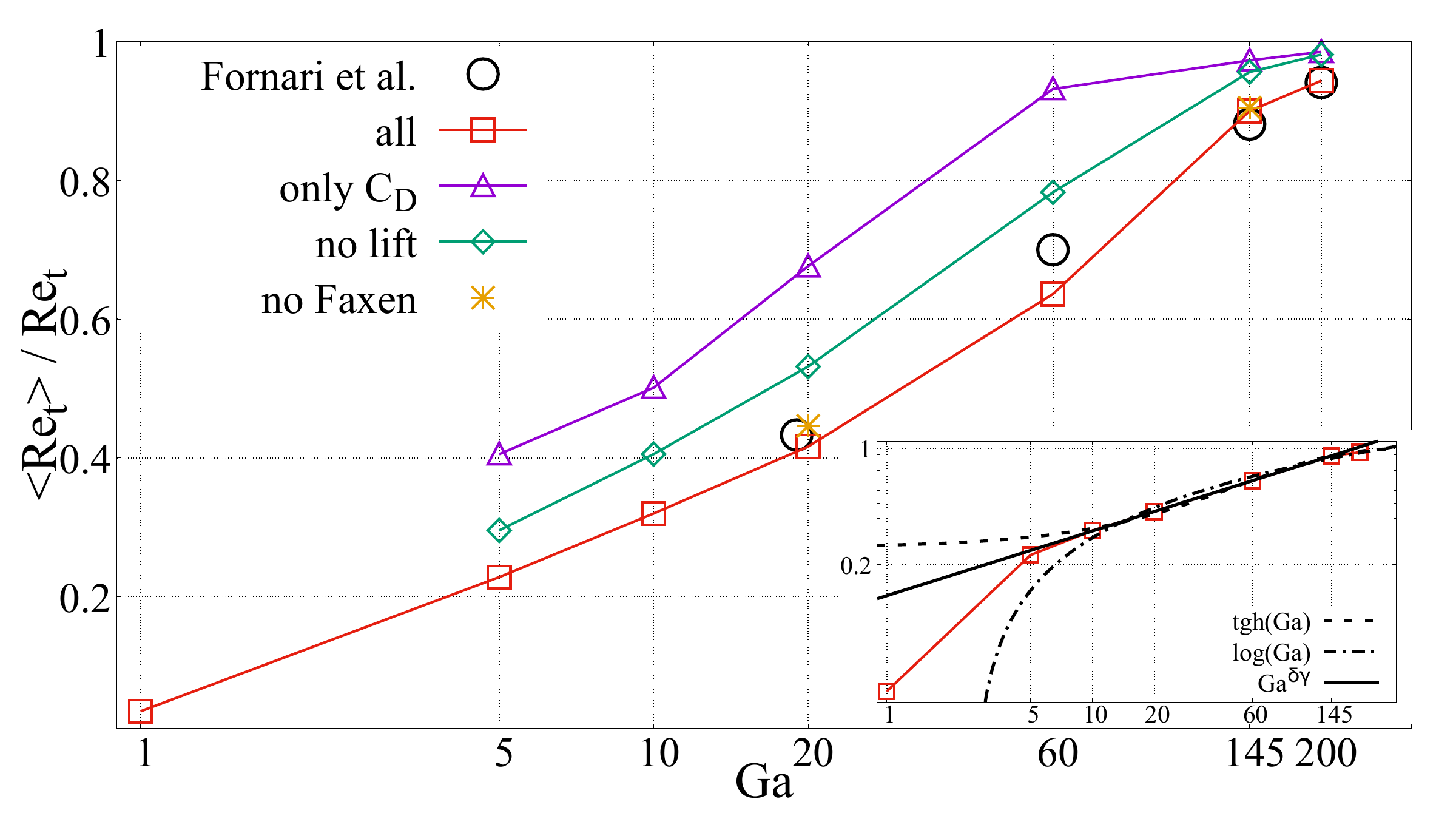}
\includegraphics[width=0.49\textwidth]{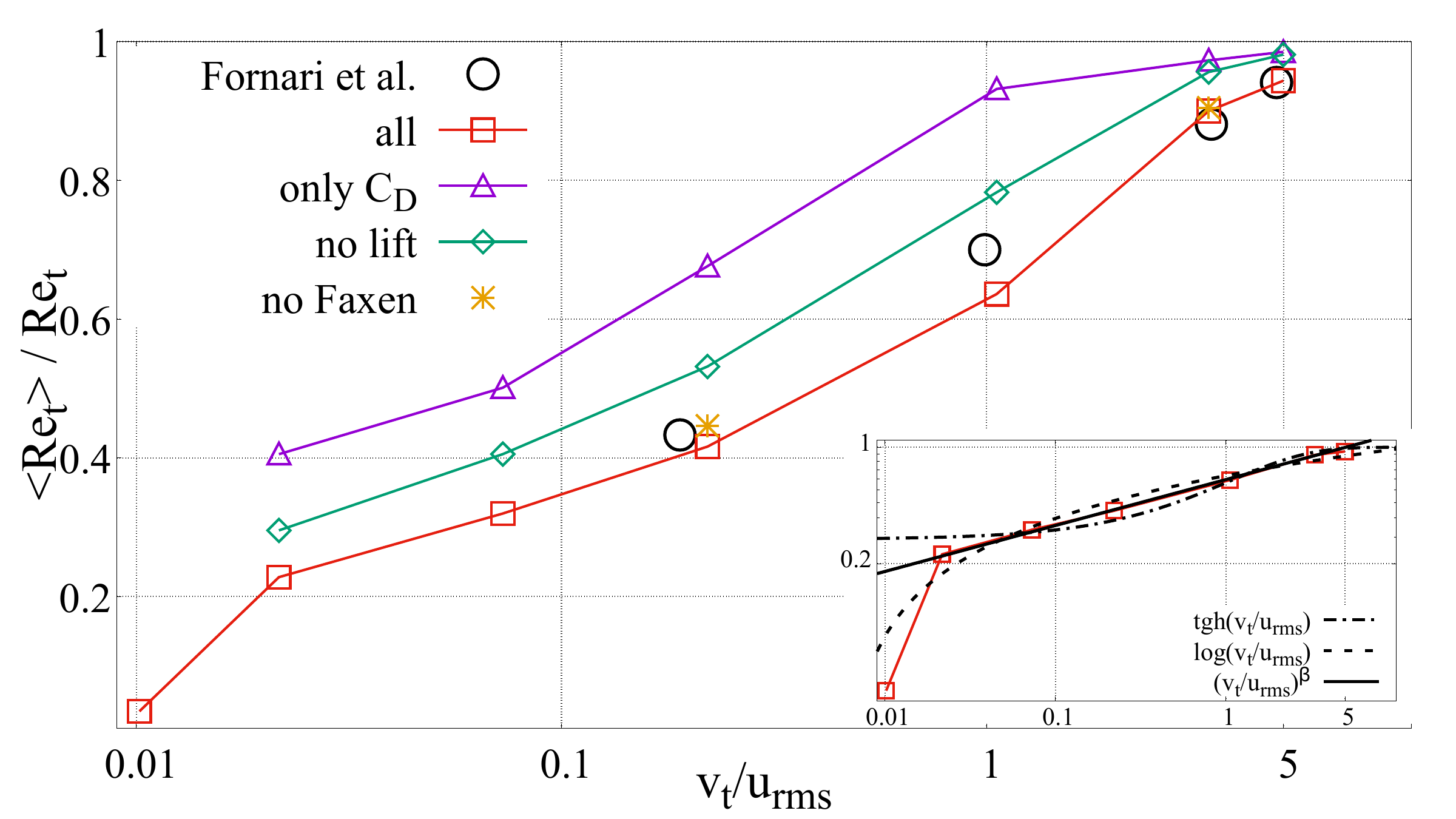}
\caption{{Normalized terminal Reynolds in the turbulent case versus Galileo number (left)
and versus the normalized terminal velocity of an isolated particle in a quiescent fluid 
$v_t/u_{rms}$ (right). The normalizing factor is $Re_t$, i.e. the 
terminal Reynolds number of a single particle in a quiescent fluid. 
Note that with the present definitions, $\lra{Re_t}/Re_t=\lra{v_t}/v_t$. 
Present data (color symbols) are compared against reference PRS 
data by Fornari et al.~\cite{fornari2016reduced} (black symbols);
Both insets highlight possible fitting scaling laws in the intermediate range using
a hyperbolic tangent, a logarithmic function, and a power law. The fitted scaling exponents are
$\delta \gamma \simeq 0.4$ (left) and $\beta \simeq 0.28$ (right).
The different color of the symbols labels the different types of simulations where we have intentionally
neglected specific terms in equation \eqref{eqn:prtcls2}, namely
\textcolor{red}{``all $\square$''}, all the forces are included,
\textcolor{Green}{``no lift $\diamond$''}, the lift force is excluded,
\textcolor{Violet}{``only $C_D$ $\Delta$''}, only the nonlinear drag is considered,
\textcolor{orange}{``no Faxen $*$''}, the Faxen correction is excluded.
}}
\label{fig:fig1}
\end{figure}

When the particles settle in a turbulent environment, the settling velocity is modified by the interaction with the carrier multi-scale flow. In the turbulent case, the terminal velocity is denoted as $\langle v_t \rangle$ where the angular bracket stands for ensemble average~\cite{pope2000turbulent}.
Correspondingly, it is possible to define a turbulent terminal Reynolds number as $\langle Re_t \rangle=\langle v_t \rangle d/\nu$. In general, in the turbulent case, data must be measured in actual experiments or simulations. 

Figure \ref{fig:fig1} shows the scaling of the terminal Reynolds number in the turbulent case $\langle Re_t \rangle$ against the Galileo number. We directly compare with the data from particle-resolved simulations (PRS) by Fornari \emph{et al.}~\cite{fornari2016sedimentation,fornari2016reduced}, which are to be thought of without any approximation and, therefore, tantamount to experiments. 
Our simulations show good agreement with the particle-resolved numerical experiments in the whole range of parameters, with a small but noticeable difference at $Ga=60$.
The turbulent settling velocity $\langle v_t \rangle$ is smaller than the settling velocity of an isolated particle in a quiescent fluid $v_t$ in all cases as in the PRS. 
It is worth noting that with the present definitions  $\lra{Re_t}/Re_t=\lra{v_t}/v_t$. 
The plot presents a nontrivial behavior: in the limit of large $Ga$, $\langle Re_t \rangle / Re_t \to 1$. 
This means that the effects of turbulence are almost irrelevant here. We observe, however, that a 
small reduction is measured even at $Ga=200$. In the opposite
range of small $Ga$, the ratio $\langle Re_t \rangle / Re_t$ decreases up to $10^{-1}$ for $Ga=1$.

Some comments are in order: $i)$ The effect of turbulence is particularly relevant in the intermediate
range of the Galileo number, where the average settling velocity is reduced by the particle-to-fluid 
interaction when compared to the quiescent case;  $ii)$ the settling velocity decreases because of the 
effect of turbulence which hinders the motion of the particles along the direction of the gravity force.

Let us discuss the plot in terms of scaling theory~\cite{barenblatt1996scaling}. 
On a dimensional ground, one expects that $\langle Re_t \rangle /{Re_t}=\Phi(Ga)$,
at fixed $ d/\eta$ and $Re_\lambda$. In our conditions, $Ga$ and $\rho_p/\rho_f$ are 
equivalent since $Ga$ is adjusted via the particle-to-fluid density ratio.
The simulations show that for large $Ga$ we have $\Phi(Ga) \to \Phi_\infty =1$ whereas
for small $Ga$ we have $\Phi(Ga) \to \Phi_0 \simeq 0$.  It follows that in the regime of 
large Galileo, the turbulent terminal velocity can be predicted using the settling velocity in 
the quiescent case. For large Galileo, particles are insensitive to turbulence and 
 behave like in the quiescent case. In contrast, in the intermediate regime 
$5 < Ga < 100$, $\Phi(Ga)$ exhibits a non-trivial behavior solely due to
the interaction between the particles and the background turbulent flow since
inter-particle interactions and two-way coupling effects are absent in our simulations.
Still, the data plotted in log-log scale, see the insets of fig \ref{fig:fig1}, show that a power law can reasonably model the scaling of $\Phi(Ga)$, as $\Phi(Ga) \propto Ga^{\delta \gamma}$ with 
$\delta \gamma \simeq 0.4$ from the best fit. Numerical evidence thus suggests that 
an intermediate scaling manifests due to the effect of turbulent fluctuations, which cannot be obtained by dimensional analysis~\cite{barenblatt1996scaling,barenblatt2003scaling}.
Indeed, in the asymptotic limits $Ga \to \infty$, we have seen that complete similarity 
is retrieved. In the intermediate range,  a correction to the scaling law $Re_t(Ga)$ of the quiescent case 
should be accounted for in the evaluation of the terminal velocity in the turbulent case. 
The exponent $\delta \gamma$ must be understood as the intermittent correction to the 
prediction of the quiescent case $Re_t \propto Ga^{\gamma}$, 
that is $\langle Re_t \rangle = Re_t \Phi(Ga) \sim Ga^{\gamma+\delta \gamma}$.
Since the scaling argument is based mainly on dimensional analysis, we also tried to fit the data 
with functions different from a power law, namely a hyperbolic tangent and a logarithm.
The power law seems the most convincing overall, most notably taking into account the points at 
$Ga=5,10$, and provides an excellent agreement over almost two decades; still the logarithm and hyperbolic tangent fits are reasonable, and the power-law scaling should not be considered as definitive. 

These results reveal that two transitions toward complete similarity regimes are present at small and large $Ga$. While the parameter $ Ga$ is appropriate at small values, it does not seem the best parameter to understand the scaling of the transition at high values.
Since the key observable is the settling speed, we show in figure \ref{fig:fig1} the settling speed in the turbulent case $\langle v_t \rangle$ versus the settling speed of an isolated particle in a quiescent fluid $v_t$ normalized by the turbulent $rms$ velocity. 
We observe that the qualitative behavior is the same, but the transition to the saturation regime happens around $v_t/u_{rms}\sim 1$. This result indicates that this ratio is a more appropriate control parameter for this problem, in line with what is proposed for settling in different configurations~\cite{maxey1987gravitational,wang1993settling}. While qualitatively equivalent, the intermittent correction is quantitatively different, being the scaling exponent $\beta \simeq 0.28$, see the inset.

In figure \ref{fig:fig1}, we also analyse the effect of the different terms in the equation (\ref{eqn:prtcls2}). To do that, we show the results of the settling speed obtained when we intentionally neglect some terms, and we compare them with the results obtained using the full equation and the reference PRS data. All the terms in equation \eqref{eqn:prtcls2} play an important role when the particle-to-fluid density ratio is of order one. The main conclusions are: $i)$ results without the nonlinear correction $C_D$ in the viscous drag are completely out of range for all values of the Galileo number greater than 1, and are not shown for clarity; $ii)$ when only the nonlinear drag is considered, the curve is much different from the reference one. Notably, the settling speed is roughly the same as in the quiescent case already at  $Ga\gtrapprox 60$, emphasising that the terms neglected are responsible for a large part of the interaction between particles and turbulence; $iii)$ the lift force appears to give a relevant contribution of about the same amount at all $Ga$. In particular, the lift force contributes to reducing the settling speed; $iv)$ the Fax\'en corrections, removed for two cases at $Ga=20,145$, seems to play a marginal role.



A final comment is to conclude the discussion of figure \ref{fig:fig1}. A small difference is present between our one-way coupling simulations and the PRS ones, which is not yet visible in the figure. Indeed, in the fully-resolved simulations, the non-zero volume fraction of particles led to a hindering effect, which also changed the settling speed in the quiescent fluid~\cite{guazzelli2011physical}. The effect was of a few percent for $Ga=60$. Since the figure shows the ratio between the turbulent settling and the corresponding quiescent one, this effect is eliminated, allowing to single out the turbulence effect. We shall retrieve this point in the conclusions. 

\begin{figure}[b!]
\centering
\includegraphics[width=0.49\textwidth]{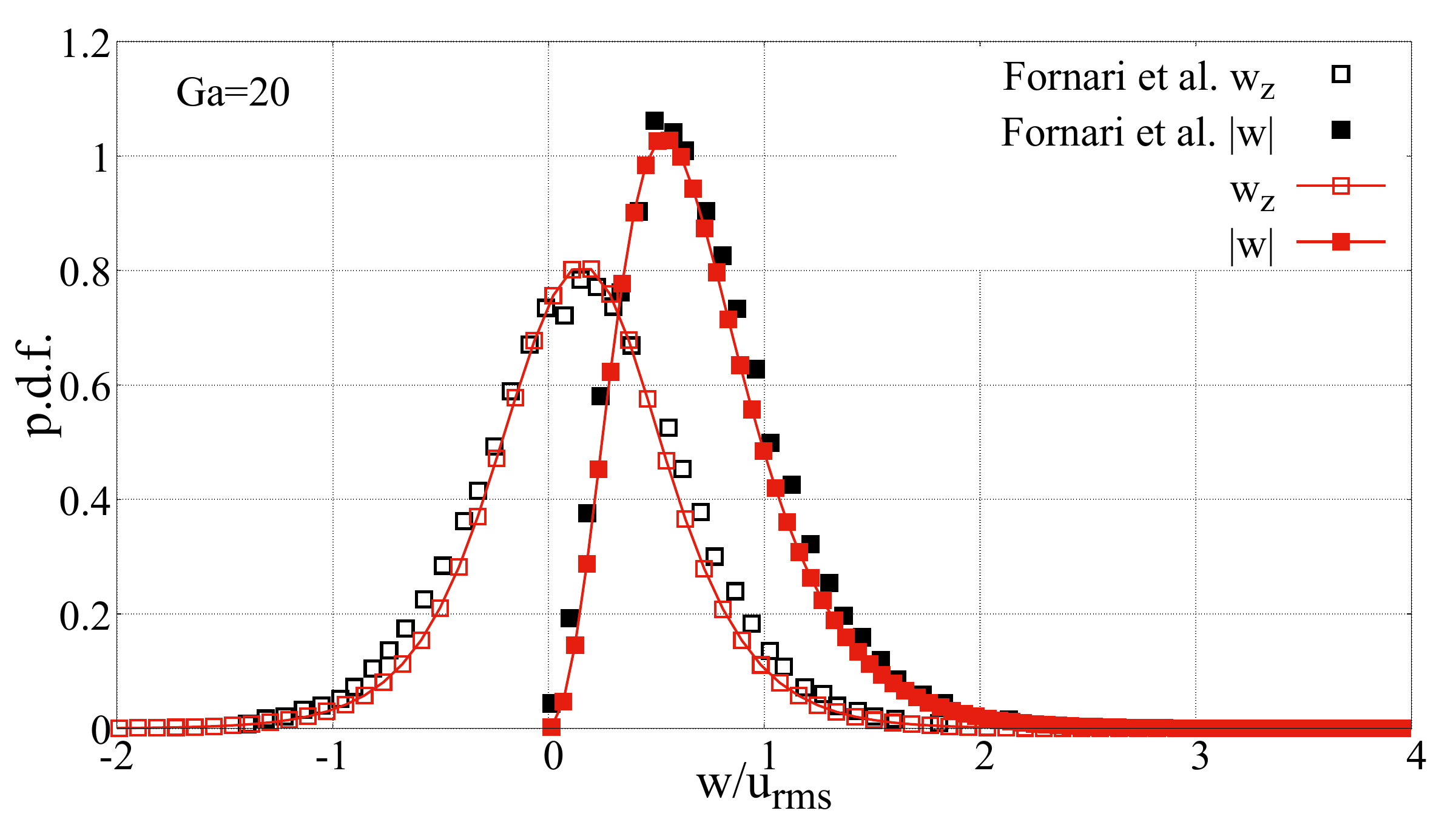}
\includegraphics[width=0.49\textwidth]{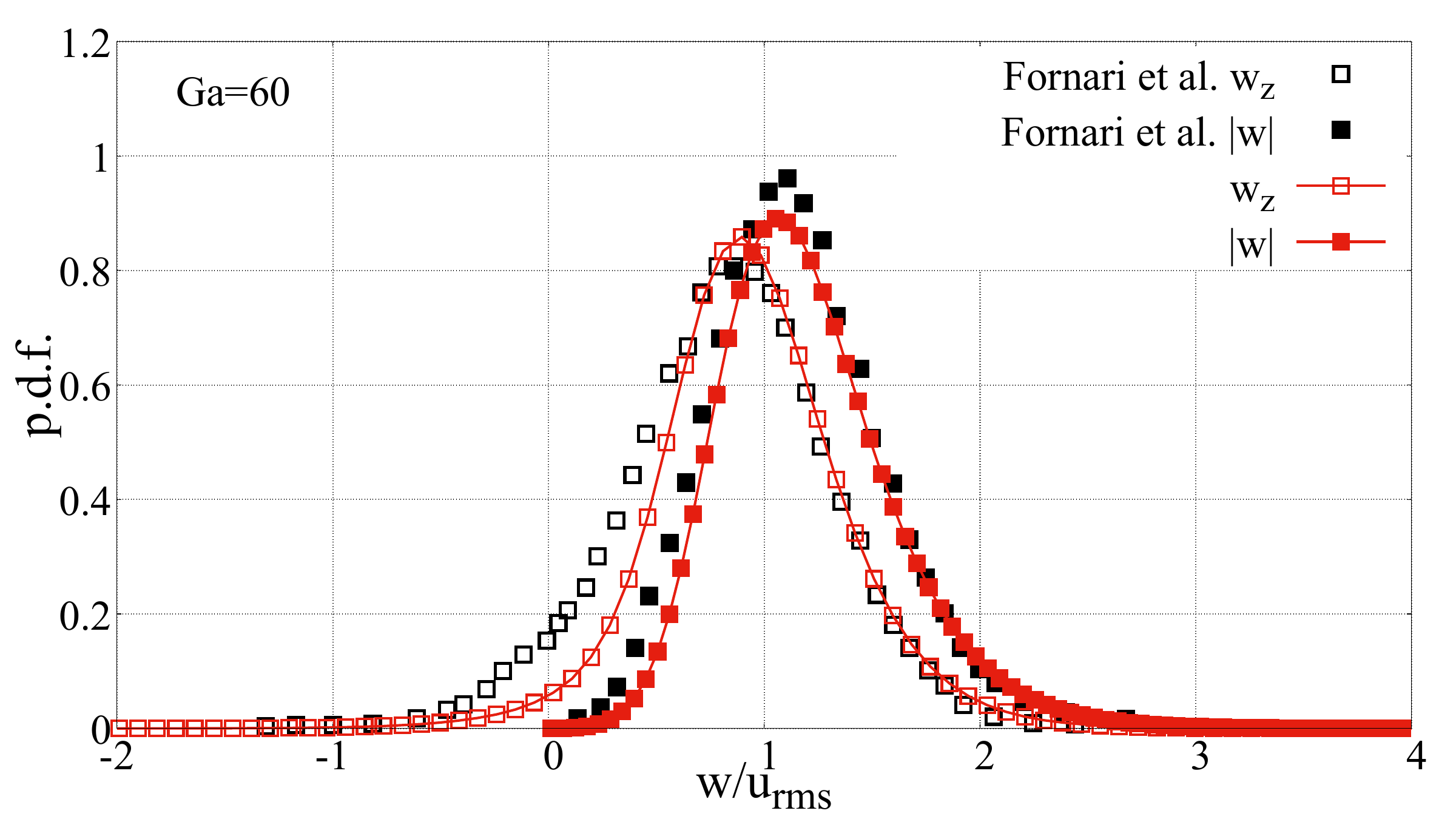}
\includegraphics[width=0.49\textwidth]{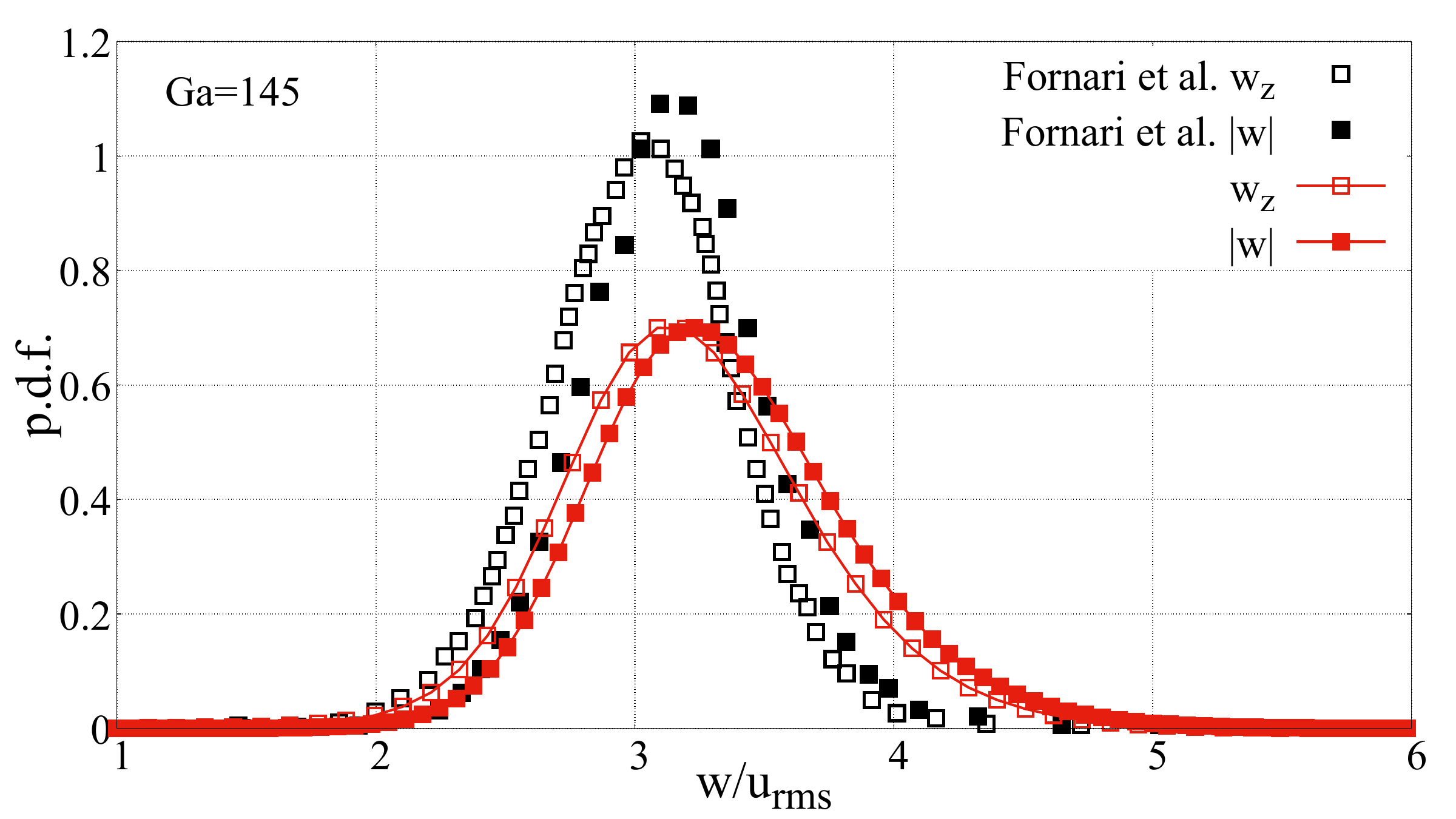}
\includegraphics[width=0.49\textwidth]{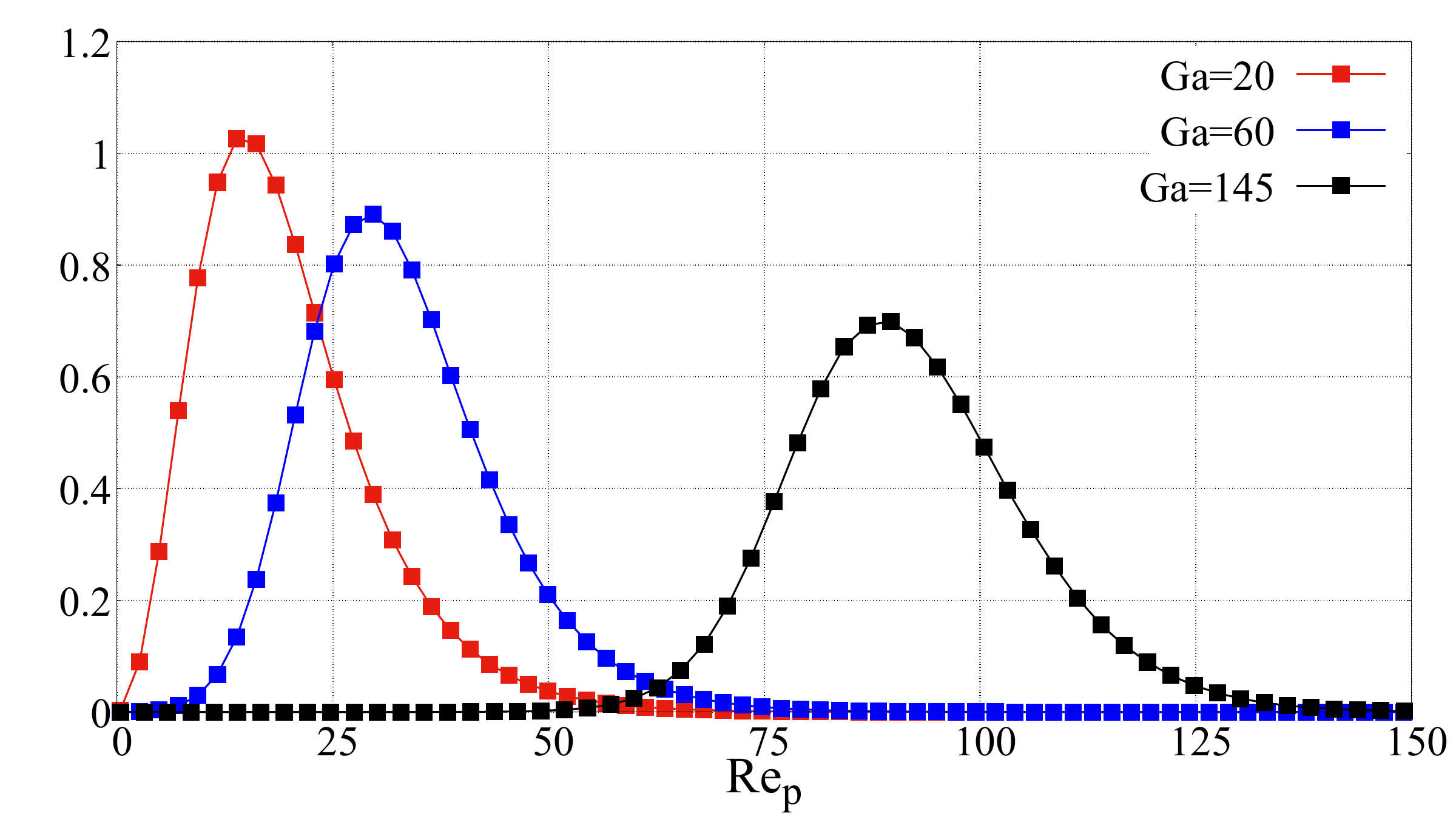}
\caption{Probability density function of the particle-to-fluid relative velocity ${\bf w}= {\bf v}_p-{\bf u}\vert_p$ . 
Both the norm (filled symbols) and the component along the direction of gravity (open symbols) are shown. 
The panels correspond to different Galileo numbers.
The bottom right panel provides the probability density function of the particle Reynolds number 
$Re_p=\vert {\bf w}\vert d/\nu$ for different Galileo numbers.
\label{fig:2}}
\end{figure}
In figure \ref{fig:2}, we show the pdf of the module of the particle-to-fluid relative velocity $|{\bf w}|$ and of the particle 
relative velocity in the direction aligned with gravity $w_z$, in comparison with the PRS data~\cite{fornari2016reduced}. Both quantities are indistinguishable at small Galileo numbers, $Ga=20$. At moderate Galileo, $Ga=60$, the agreement remains very good, with possibly some difference for the smallest values of the relative velocity, see the left tail of the pdf. 
In particular, there is a slightly larger probability of finding negative values of the velocity along 
the gravity direction, whereas the module remains essentially the same. 
Note that the case at $Ga=60$ is the one for which a slight difference is also encountered for the 
settling velocity. At higher Galileo $Ga=145$, differences are more important. While the qualitative trend is still well captured, including the mode of the distribution, the shape of the pdf in the point-wise approach differs from the PRS prediction, with a distribution more peaked around the mean in the PRS case. 
This means that the particle-resolved simulations filter the surrounding fluctuations more efficiently. 
That is not so surprising, since as proposed by~\cite{clift2005bubbles}, the Galileo number can 
be understood as a dimensionless diameter, while dimensionless terminal velocity does not depend on the particle diameter. Thus, at large Galileo numbers, finite-size effects are expected to be important in filtering small-scale velocity fluctuations, whereas the point-wise model incorporates only first-order corrections~\cite{happel2012low}. Note, however, that in all cases, the relation between the relative velocity along the gravity and its absolute value is well represented, indicating that the physics in both the horizontal plane and vertical direction is satisfactorily described in the present point-wise approach.

To conclude, the results about absolute relative velocity can be recast in terms of a pdf of the particle Reynolds number, see the bottom-right panel in figure~\ref{fig:2}. The more frequent occurrences are $Re_p\sim 10$ at $Ga=20$, and $Re_p\sim 80$ at $Ga=145$, with fluctuations indicating particles with $Re_p>100$. 
These results emphasize that a significant relative velocity is encountered with increasing Galileo number. Remarkably, despite the fairly large value of $Re_p$, our point-wise model with the nonlinear drag 
correction is capable of reasonably well reproducing in all the cases the mean particle settling velocity, 
see figure \ref{fig:fig1}.

\begin{figure}[b!]
\centering
\includegraphics[width=0.49\textwidth]{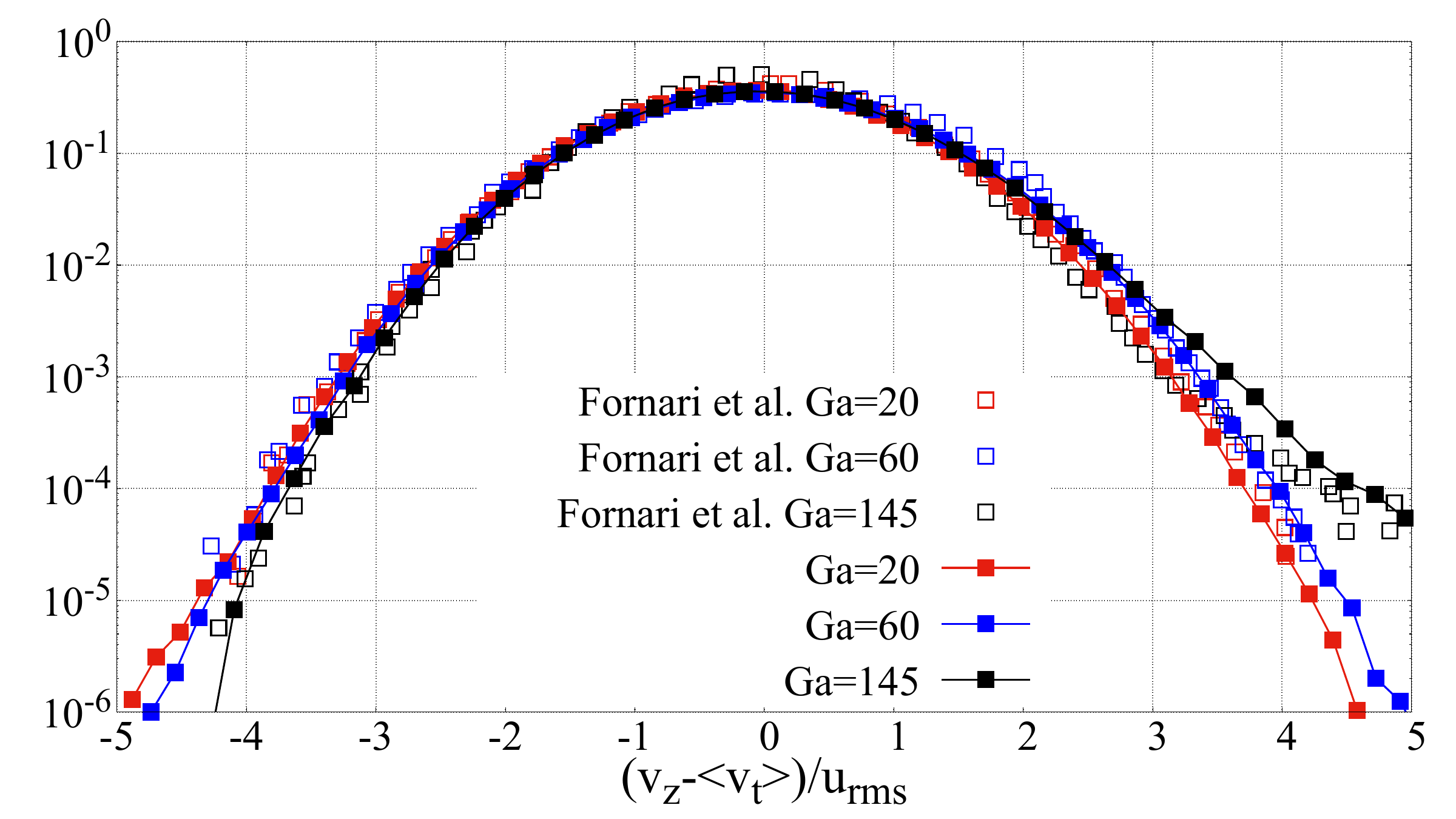}
\includegraphics[width=0.49\textwidth]{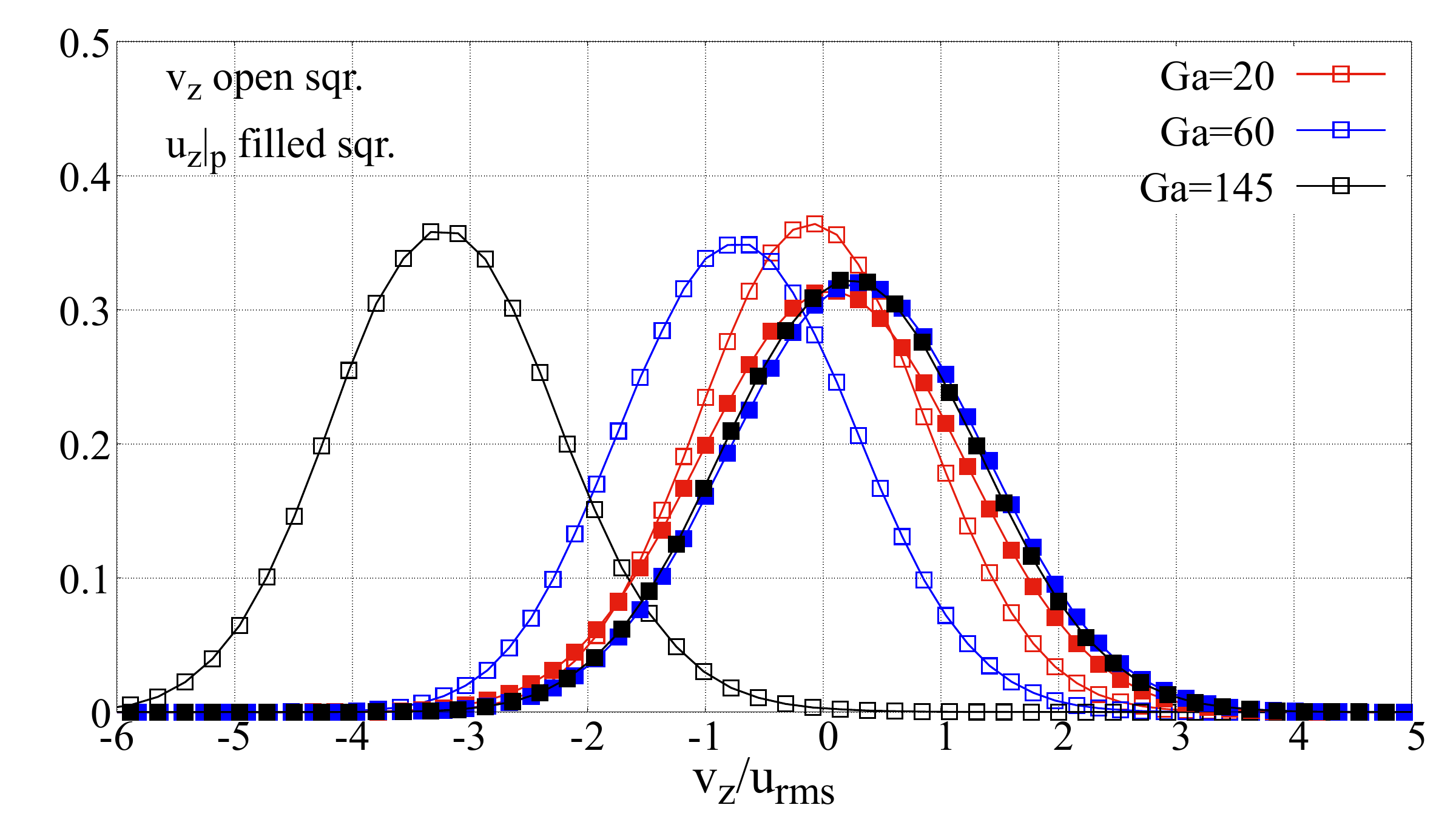}
\caption{ Left panel: probability density function of the centered vertical particle velocity at different Galileo 
numbers (filled symbols). For comparison, the PRS data by Fornari et {\em al.} are reported (open symbols). Data 
have been sampled from the figures in \cite{fornari2016reduced}.
Right panel: probability density function of the particle velocity (open symbols) and the fluid velocity seen 
by the particles (filled symbols) along the direction of gravity at different Galileo numbers.} 
\label{fig:3}
\end{figure}
The fluctuations of the particle velocity are analyzed in figure~\ref{fig:3}.
In the left panel, we show the pdf of the particle velocity along the direction of gravity. 
The distribution is centered by removing the average settling speed and normalized with the 
turbulence $rms$ velocity. It turns out that the pdf is approximately Gaussian 
in the point-wise simulations, yet with the skewness slightly increasing with Galileo. 
This effect is concentrated at rare positive events, while the negative tails remain
almost unchanged. As for the relative velocities, the agreement is excellent with the PRS at 
$Ga=20,60$. More slight differences emerge (as expected) at $Ga=145$ in the right tail of the pdf
corresponding to relatively more likely but stronger velocity fluctuations. 
In any case, considering the drastic approximations carried out in the point-wise approach, the data generally agree satisfactorily with PRS reference curves, indicating that the point-wise approach is able to reasonably reproduce the particle velocity fluctuations.

In the right panel of figure~\ref{fig:3}, we analyse the same pdf of the particle velocity along 
the direction of gravity without removing the average settling speed compared with the fluid velocity seen by the particles. As expected, the pdf is shifted toward higher velocities with increasing the Galileo number, consistently to what was observed in PRS of Fornari et {\em al.}~\cite{fornari2016reduced}. The pdf of the velocity seen by the particles (closed symbols) highlights the preferential sampling of the updraft regions of turbulence; that increases with $Ga$, as the mean becomes increasingly positive. While the preferential sampling seems to be equivalent for  $Ga\ge 60$, the impact of the settling velocity is much different because of the difference in the mean value. This evidence shows a small ``loitering" mechanism~\cite{nielsen1993turbulence} at larger Galileo.

\begin{figure}[h!]
\includegraphics[width=0.32\textwidth]{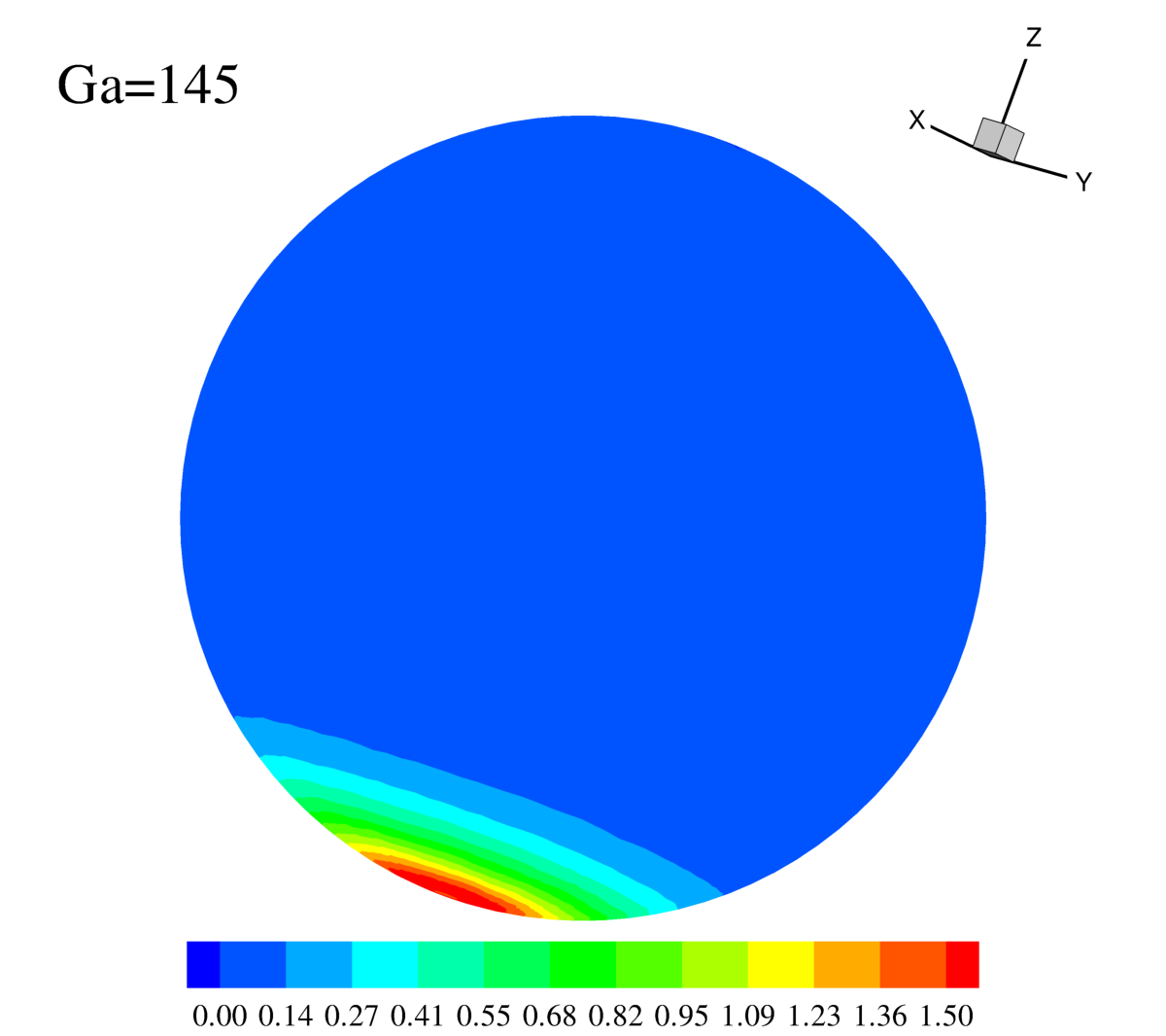}
\includegraphics[width=0.32\textwidth]{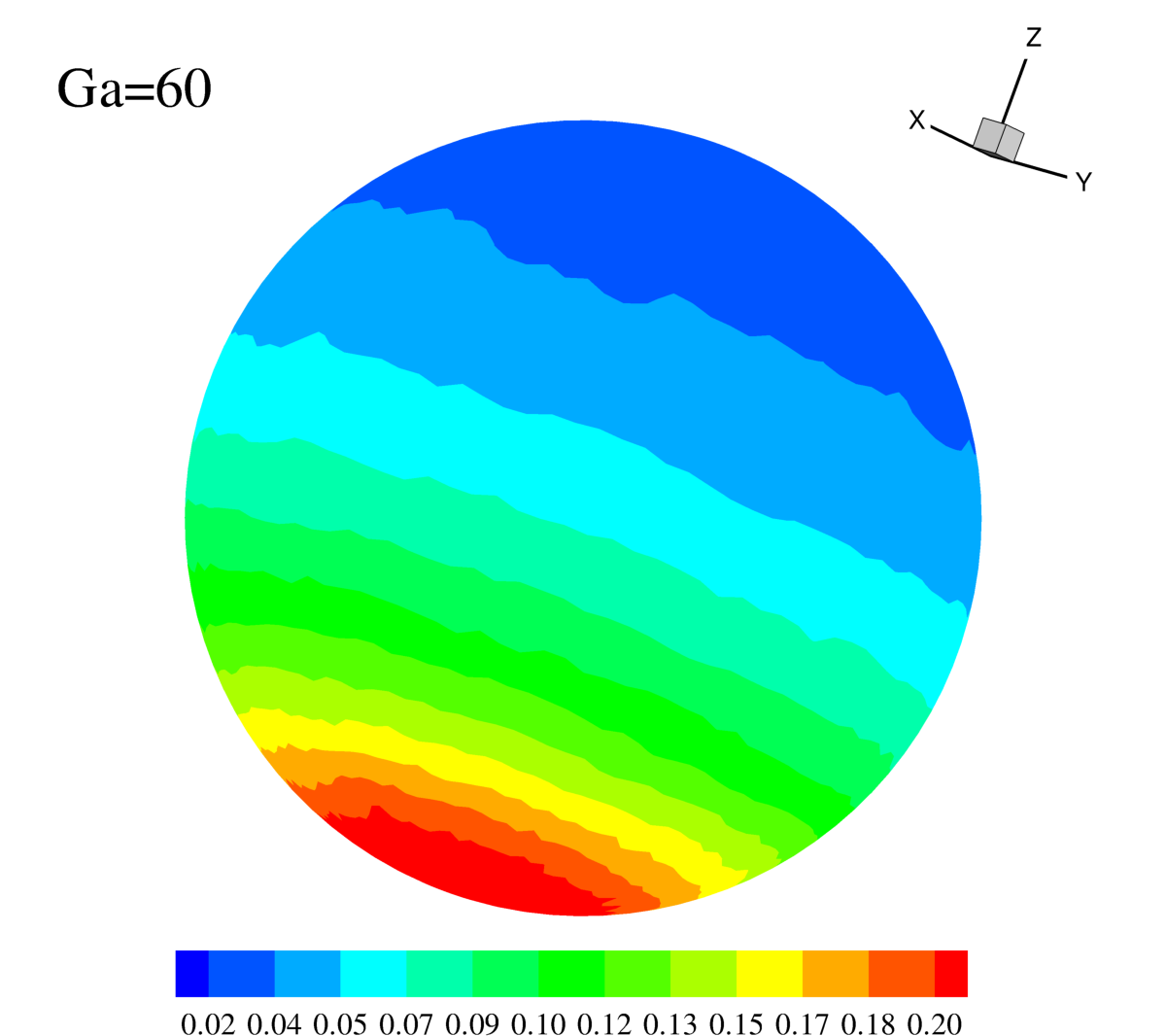}
\includegraphics[width=0.32\textwidth]{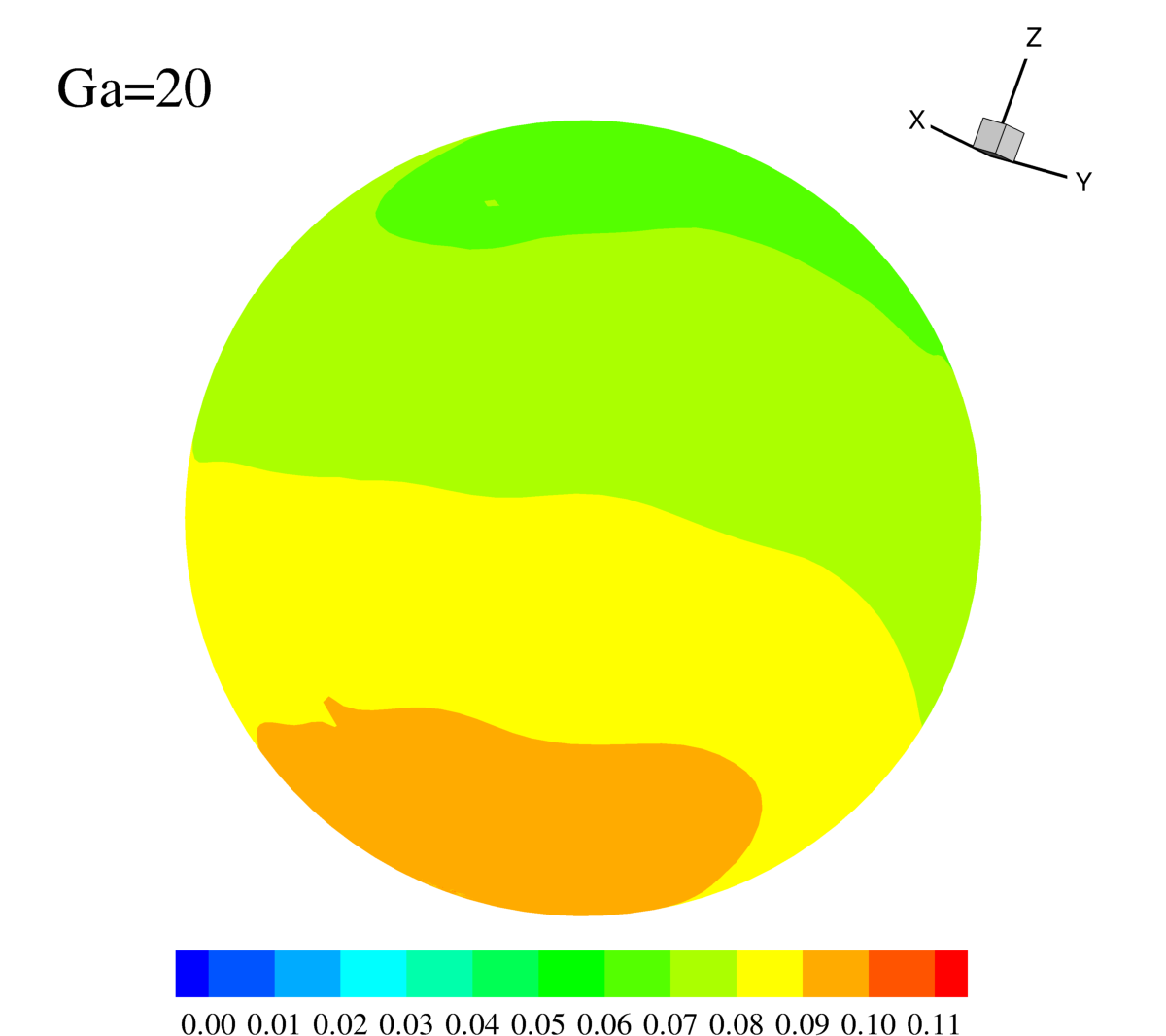}
\caption{Probability distribution function of the particle unit velocity vector direction 
represented on the unit sphere at different Galileo numbers. In the plot, the gravity is along the negative z-axes.}
\label{fig:fig4}
\end{figure}

\begin{figure}[b!]
\includegraphics[width=0.98\textwidth]{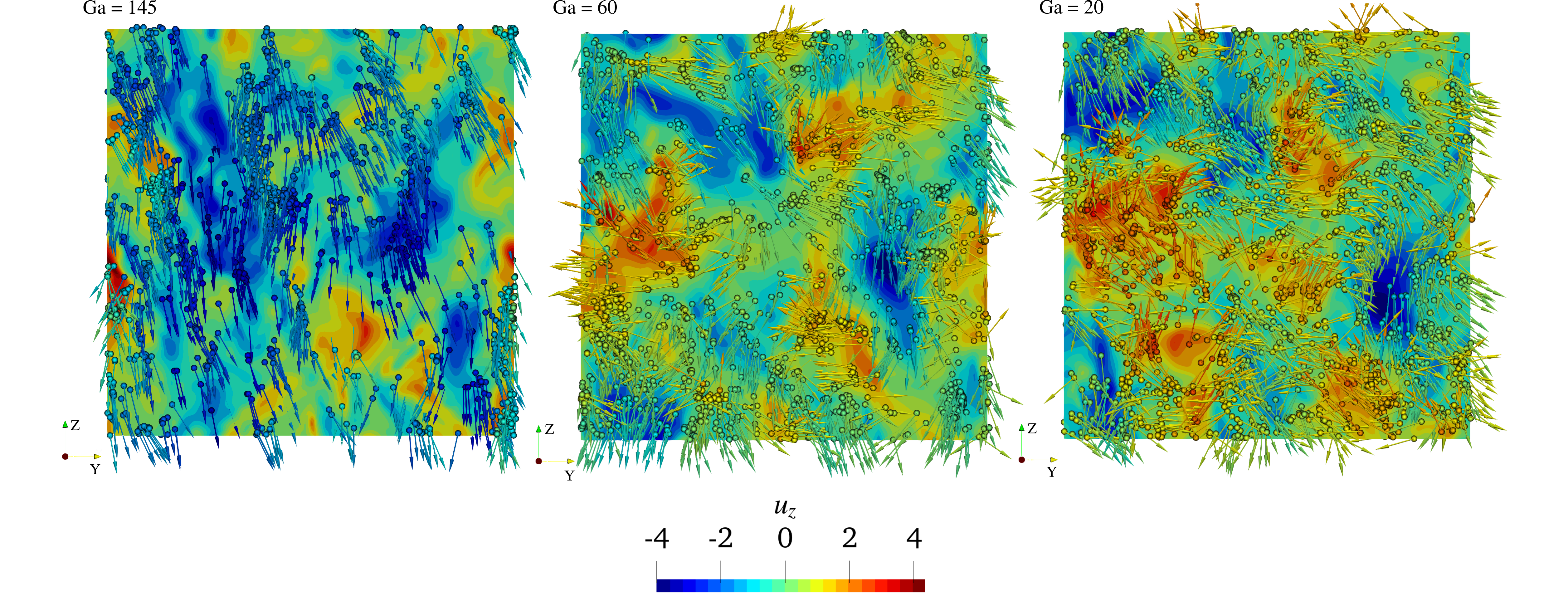}
\caption{Snapshot of the z-component of the fluid velocity field (contour plot) and particle position in a
$z-y$ plane. The gravity is along the negative $z$ direction. The vectors represent the in-plane direction of
the particle velocity and are colored with the corresponding z-component of the velocity. Cases with the same Galileo number as in figure \ref{fig:fig4}.}
\label{fig:fig4inst}
\end{figure}

These results are corroborated by the analysis shown in figure \ref{fig:fig4}. The contours on the unit sphere 
represent the probability of finding the particle velocity unit vector in a specific direction of the space. 
Here, the gravity force is in the negative z-direction, which is an axis of symmetry for the problem. 
The plot conveys a geometrical view of the settling phenomena since it highlights at a glance the 
most preferred direction in space that, on average, the particles follow during their settling.
At large Galileo ($Ga=145$), particles mostly fall along the gravity direction with weak probabilities to
experience lateral velocities. There is, therefore, very little difference with particles settling in quiescent fluid. 
At intermediate Galileo ($Ga=60$), while particle velocity remains mostly oriented toward the gravity direction, 
they experience large contributions from lateral velocities, and a fairly large range of directions is explored.  
At small Galileo ($Ga=20$), the particle velocity vector is likely to point in an even broader range of directions with
a finite probability of updraft events. This result shows that turbulence can cause particles to remain floating, 
with a corresponding reduction of their average settling velocity.
The above statistical characterization of the settling can
be visually appreciated by the inspection of the instantaneous flow field and particle positions and velocities.
Figure \ref{fig:fig4inst} visualises the instantaneous z-component of the velocity field in a plane containing the
gravity and the corresponding particle position. The vectors represent the direction of the in-plane particle velocity 
colored with their z-component. From the inspection of the instantaneous field, at $Ga=145$, the particle velocities are
mainly aligned in the negative z-direction. Indeed, the particles have velocities that substantially differ from the 
background fluid velocity, meaning that the main driving effect is the gravity force. As the Galileo number
is decreased, cases at $Ga=60$ and $Ga=20$, the effect of turbulent fluctuations becomes more and more important.
The particle velocities explore a wider range of spatial directions and are more similar to the local 
(i.e. in the particle position) fluid velocity.

\begin{figure}[b!]
\begin{center}
\includegraphics[width=0.32\textwidth]{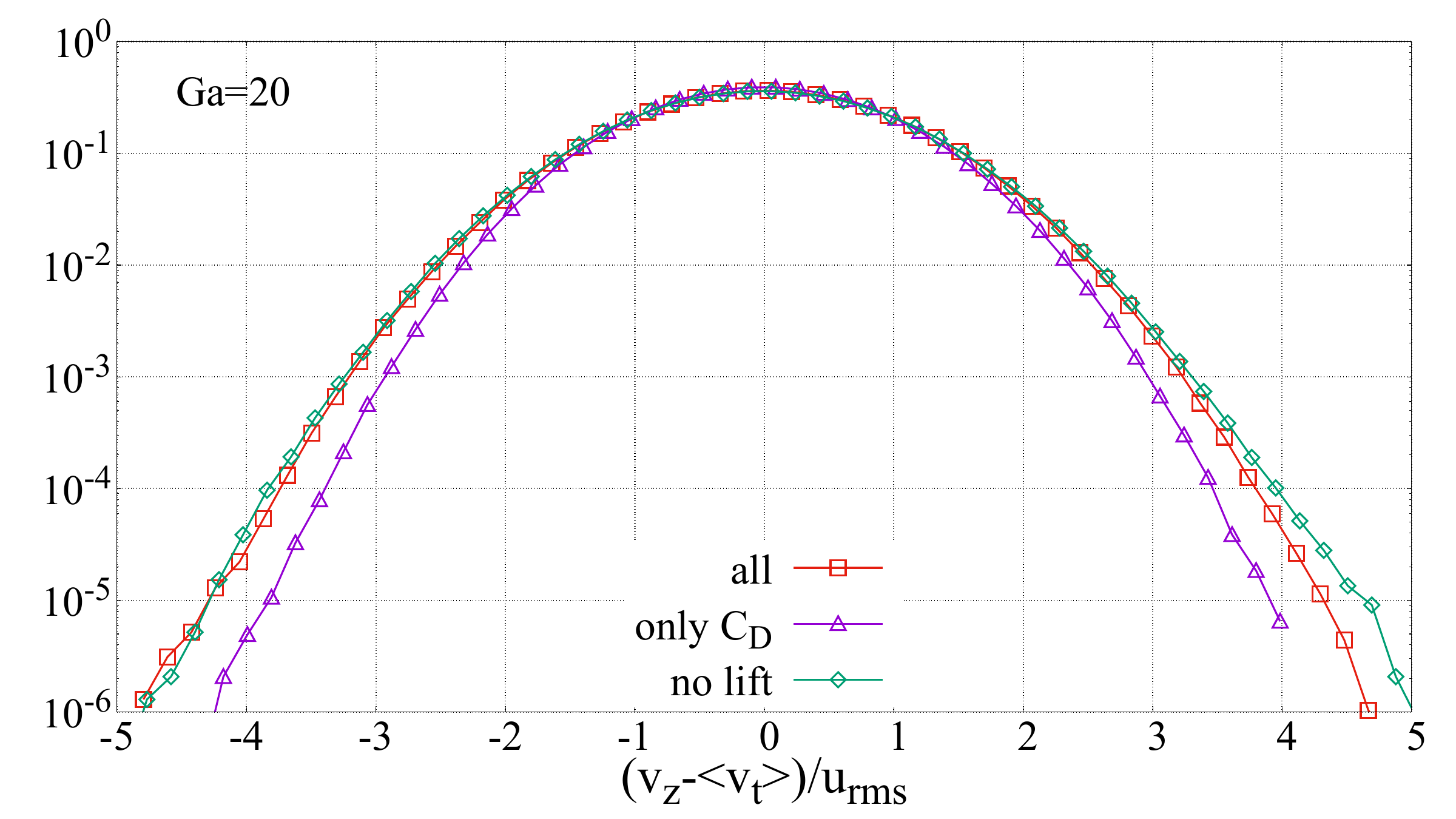}
\includegraphics[width=0.32\textwidth]{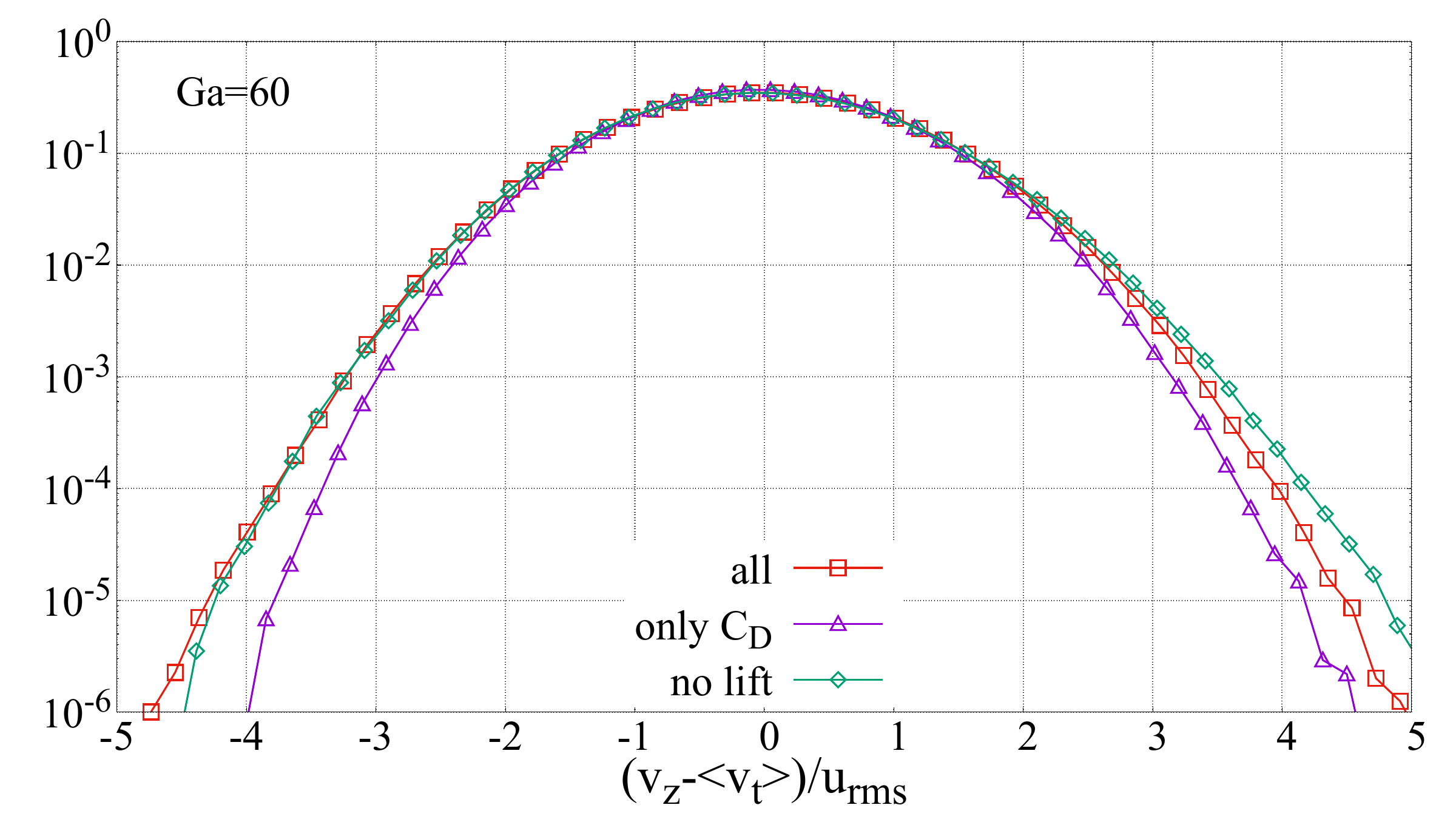}
\includegraphics[width=0.32\textwidth]{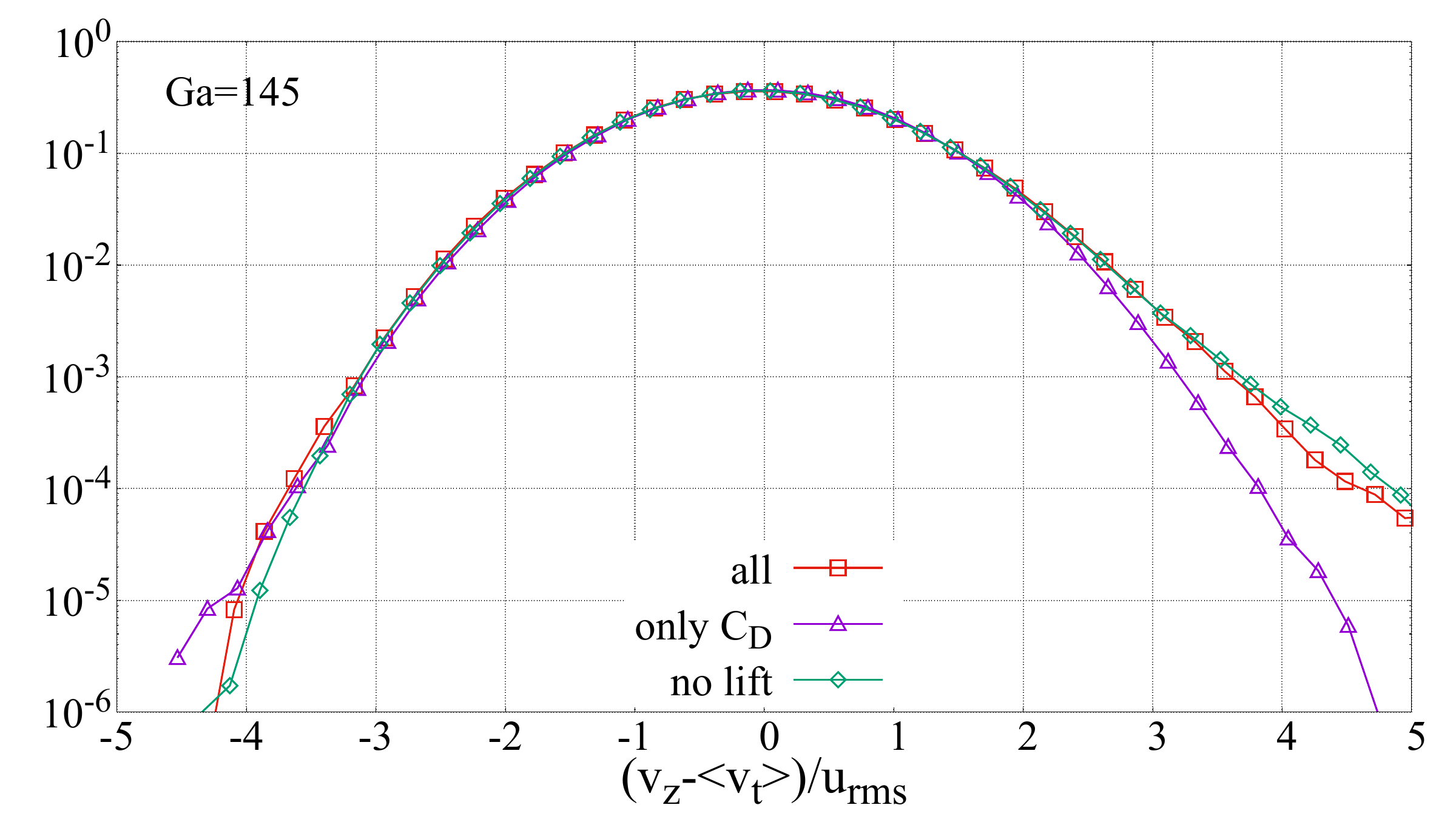}
\caption{Probability density function of the centered particle velocity along the gravity direction at different 
Galileo numbers. Symbols correspond to simulations where intentionally the lift force was neglected and where
only the nonlinear viscous drag was considered.
\label{fig:5}}
\end{center}
\end{figure}

To conclude the analysis, in figure~\ref{fig:5}, we show how the lift force term and the unsteady term in equation \eqref{eqn:prtcls2} affect the pdf of the particle velocities.
The lift force is negligible at a high Galileo number. At a moderate Galileo number, it has a small impact except in the tails of positive velocities, where the lift decreases the probability of extreme velocities. On the other hand, if the unsteady terms are also switched off (data labeled ``only $C_D$''), there is a stronger impact on the pdf, with a measurable shrinking of the variance while the global shape is conserved. In conclusion, the non-linear drag force is the key to capturing the global dynamics, 
yet all terms are important to achieve quantitative predictions.

\section{Conclusions}

In this work, we have revisited the important problem of particles settling under gravity in a turbulent flow with a rough model of point-wise particle and one-way coupled direct numerical simulations. We have chosen the physical parameters considered before in the framework of particle-resolved simulations at semi-dilute conditions, i.e. particles larger than the Kolmogorov scale but with a particle-to-fluid density ratio of order one.
While dilute heavy (i.e. large density ratios) particle-laden flows have been comprehensively 
investigated,  few studies have considered almost neutrally buoyant particles, and none through the one-way coupling approach (to the best of our knowledge). 
Interestingly, the results tell a different story concerning what could be inferred from the previous investigations.

We have found that the turbulent settling velocity decreases with a decreasing Galileo number, which is in good agreement with the reference particle-resolved data. Therefore, the first conclusion is that two-way coupling, finite-size effects, and inter-particle interactions appear negligible to get the correct average turbulent settling speed in the range of parameters considered here.
It is important to note that we have considered some finite-size effects through Fax\`en corrections. However, their contribution is found to be small while it is
of crucial importance to consider unsteady and lift force terms.
This result is of practical importance since point-wise simulations may also be used 
as an effective tool for realistic flows, whereas resolved simulations cannot.

In the large $Ga$ limit, the settling velocity of an isolated particle in a quiescent fluid is retrieved since particles become insensitive to turbulent fluctuations.
In the small Galileo limit, results suggest that turbulence hinders settling.
In the intermediate range of Galileo numbers, a power-law reasonably models the scaling of the normalised settling velocity over almost two decades. This power-law is consistent with relating the turbulent reduction in settling speed to an intermittent correction to the quiescent behavior. In other words, there is an incomplete similarity in this intermediate range of Galileo numbers.

We have also discussed the relevance of the different terms arising in Basset-Boussinesq-Oseen equations, i.e. non-linear drag, pressure gradient, added-mass, and lift force. 
The rigorous form of these equations~\cite{gatignol1983faxen} is valid only at $Re_p\ll 1$.
We have therefore added a standard nonlinear correction to the drag force since, as shown by the simulations, the terminal Reynolds number is much higher than one in almost all cases. This is a subtle point since the nonlinear correction has been heuristically proposed for heavy particles, and it has not been fully assessed for $\rho_p / \rho_f \approx 1$ as used here. 
In the original derivation of~\cite{gatignol1983faxen}, the lift force is absent, and its precise derivation is still a matter of ongoing research~\cite{candelier2008time,candelier2019time}. In the present work, we have used the standard Auton's formula, but a more exhaustive analysis of this issue is an interesting perspective for the future.

The analysis of the contributions of the different terms indicates that point-wise simulations with 
all these terms can correctly reproduce the physics pointed out by fully-resolved simulations, most notably at small and moderate Galileo numbers. In particular, the nonlinear drag is always the most important contribution. In the high Galileo number regime, a small loitering effect is found.
The lift term is found to contribute to the range of parameters considered. In particular, it may 
help to reproduce some of the unsteady contributions which are found to increase for 
$Ga\simeq 60$ in resolved simulations. That is in line with the past analysis of this force~\cite{candelier2008time}.


In fully resolved simulations, a hindering effect may be recorded~\cite{guazzelli2011physical}, which cannot be present in one-way simulations. Still, for the volume fraction considered here, the effect should be small~\cite{yin2007hindered,uhlmann2014sedimentation}.
Furthermore, the normalized settling velocity is not impacted by this effect, and since the analytical formula for the hindering is available~\cite{yin2007hindered}, that should not be a drawback from a practical point of view, even at larger concentrations. 
The volume fraction considered in the fully-resolved simulations $\Phi_V=0.5\%$ has been found to be too small to trigger appreciable coupling effects, which seems in line with 
the results known for the heavy particle regime~\cite{monchaux2017settling}.
The absence of a substantial finite-size effect of the particle is less intuitive since the diameter is much larger than the Kolmogorov scale. However, in the present configuration, only the Galileo number is the relevant parameter. Indeed, some discrepancy in velocity fluctuations between present simulations and fully-resolved ones are found only at very large Galileo, where, however, the impact of the turbulence is in any case negligible.

In conclusion, the present results are remarkable since they show that the settling velocity is well captured at all Galileo numbers for particles larger than the Kolmogorov scale at density ratio order one when all the terms in the Basset-Boussinesq-Oseen equations are considered adding the heuristic contribution of non-linear drag force and lift force. For a volume fraction of a few percent, it is shown that a one-way coupling regime is sufficient to capture the relevant physics of the settling. This makes point-wise simulations a viable option for applications where particle-resolved simulations are not yet feasible.

\section{Acknowledgements}
Fruitful discussions with E. Guazzelli are kindly acknowledged. We acknowledge the CINECA award under the ISCRA initiative, Iscra B number HP10B0F5V3. This work received financial support from the Sapienza 2021 Funding Scheme, project no. RG12117A66DC803E.

\bibliographystyle{unsrt}
\bibliography{biblio}
\end{document}